\def\gsim{\;\lower4pt\hbox{${\buildrel\displaystyle >\over\sim}$}\,}
\def\lsim{\;\lower4pt\hbox{${\buildrel\displaystyle <\over\sim}$}\,}
\def\mach{{\cal M}}
\def\ener{{\overline{\cal E}}\,}
\def\em{{\rm em}}
\def\FLASH{{\sc flash}}

\documentclass{aa}  
\usepackage{graphicx}
\usepackage{natbib}
\usepackage[english]{babel}
\usepackage{txfonts}
\begin{document}
   \title{Crushing of interstellar gas clouds in supernova remnants}

   \subtitle{II. X-ray emission}

   \author{S. Orlando\inst{1},
           F. Bocchino\inst{1},
           G. Peres\inst{2,1},
           F. Reale\inst{2,1},
           T. Plewa\inst{3,4},
          \and
           R. Rosner\inst{3,4,5}
          }

   \offprints{S. Orlando,\\ e-mail: orlando@astropa.inaf.it}

   \institute{INAF - Osservatorio Astronomico di Palermo ``G.S.
              Vaiana'', Piazza del Parlamento 1, I-90134 Palermo, Italy
         \and
              Dip. di Scienze Fisiche \& Astronomiche, Univ. di
              Palermo, Piazza del Parlamento 1, I-90134 Palermo,
              Italy
         \and
              Dept. of Astronomy and Astrophysics,
              University of Chicago,
              5640 S. Ellis Avenue, Chicago, IL 60637, USA
         \and
              Center for Astrophysical Thermonuclear Flashes,
              University of Chicago,
              5640 S. Ellis Avenue, Chicago, IL 60637, USA
         \and
              Argonne National Laboratory, 9700 South Cass Avenue,
              Argonne, IL 60439-4844, USA
             }

   \date{Received \quad\quad\quad ; accepted \quad\quad\quad }

   \authorrunning{S. Orlando et al.}
   \titlerunning{Crushing of interstellar gas clouds in supernova
                 remnants. II.}

 
  \abstract
   {X-ray observations of evolved supernova remnants (e.g. the Cygnus
    loop and the Vela SNRs) reveal emission originating from the
    interaction of shock waves with small interstellar gas clouds.}
   {We study and discuss the time-dependent X-ray emission predicted by
    hydrodynamic modeling of the interaction of a SNR shock wave with
    an interstellar gas cloud. The scope includes: 1) to study the
    correspondence between modeled and X-ray emitting structures, 2)
    to explore two different physical regimes in which either thermal
    conduction or radiative cooling plays a dominant role, and 3) to
    investigate the effects of the physical processes at work on the
    emission of the shocked cloud in the two different regimes.}
   {We use a detailed hydrodynamic model, including thermal conduction
    and radiation, and explore two cases characterized by different Mach
    numbers of the primary shock: $\mach = 30$ (post-shock temperature
    $T_{\rm psh} \approx 1.7$ MK) in which the cloud dynamics is dominated
    by radiative cooling and $\mach = 50$ ($T_{\rm psh} \approx 4.7$
    MK) dominated by thermal conduction. From the simulations, we
    synthesize the expected X-ray emission, using available spectral
    codes.}
   {The morphology of the X-ray emitting structures is significantly
    different from that of the flow structures originating from the
    shock-cloud interaction. The hydrodynamic instabilities are never
    clearly visible in the X-ray band. Shocked clouds are preferentially
    visible during the early phases of their evolution. Thermal
    conduction and radiative cooling lead to two different phases of
    the shocked cloud: a cold cooling dominated core emitting at low
    energies and a hot thermally conducting corona emitting in the X-ray
    band. The thermal conduction makes the X-ray image of the cloud
    smaller, more diffuse, and shorter-lived than that observed when
    thermal conduction is neglected.}
   {}


   \maketitle
%
\section{Introduction}
\label{intro}

This paper is part of a series devoted to study the interaction of a shock
wave of an evolved supernova remnant (SNR) with a small interstellar
gas cloud (like the ones observed, for instance, in the Cygnus loop or
in the Vela SNR) through detailed and extensive numerical modeling. The
project aims at overcoming some of the limitations found in the previous
analogous studies (for instance, taking into account simultaneously
important physical effects such as heat flux and radiative losses)
and crucial for the accurate interpretation of the high resolution
multi-wavelength observations of middle-aged SNR shell available with
the last-generation observatories.

In a previous paper (\citealt{2005A&A...444..505O}; hereafter Paper I),
we have modeled in detail the shock-cloud interaction with hydrodynamic
simulations including the effects of thermal conduction and radiative
losses from an optically thin plasma. We have investigated the interplay
of the latter two processes on the cloud evolution and on the mass
and energy exchange between the cloud and the surrounding medium, by
exploring two different physical regimes in which one of the two processes
is dominant. We have found that, when the radiative losses are dominant,
the shocked cloud fragments into cold, dense, and compact filaments
surrounded by a hot corona gradually ablated by the thermal conduction;
to the contrary, when the thermal conduction is dominant, the shocked
cloud evaporates in a few dynamical time-scales. In both cases we have
found that the thermal conduction is very effective in suppressing the
hydrodynamic instabilities that would develop at the cloud boundaries.

In this paper, we study and discuss the time-dependent X-ray emission
predicted by the modeling mentioned above. Several authors have
investigated the emission from SNRs in different spectral bands,
adopting global SNR models including the effects of radiative cooling,
evaporating clouds, large scale gradients for the ISM density etc. (e.g
\citealt{1981ApJ...247..908C}, \citealt{1991ApJ...373..543W},
\citealt{1999ApJ...524..179C}, \citealt{1999ApJ...524..192S},
\citealt{1999A&A...344..295H}, \citealt{2001A&A...371..267P},
\citealt{2004ApJ...601..885V}). On the other hand, the X-ray emission
originating from model shocked interstellar clouds has not been
investigated yet in detail, despite it is potentially important for
the energy budget of the shocked ISM and for the interpretation of the
observations.

Here we synthesize from the numerical simulations described in Paper I
the X-ray emission expected from the shock-cloud interaction. 
Our scope includes: 1) to link modeled to X-ray emitting structures,
2) to investigate the emission of the shocked cloud in two different
physical regimes in which either thermal conduction or radiative cooling
is dominant, and 3) to investigate the effects of thermal conduction
and radiation on the emission of the shocked cloud.

The paper is structured as follows: in Sect. \ref{sec2} we briefly
summarize the model of the shock-cloud collision and outline the method
to synthesize the X-ray emission from the numerical simulations; in
Sect. \ref{sec4} we discuss the results; and finally in Sect. \ref{sec5}
we draw our conclusions.

\section{The modeling}
\label{sec2}

\subsection{Hydrodynamic simulations}

In this section, we summarize the model of the shock-cloud collision.
We refer the reader to Paper I for more details.

The model describes the impact of a planar shock front onto an isolated
gas cloud. The cloud before the impact is assumed to be spherical with
radius $r_{\rm cl} = 1$~pc, small compared to the curvature radius of the
SNR shock\footnote{In the case of a small cloud, the SNR does not evolve
significantly during the shock-cloud interaction, and the assumption of
a planar shock is justified (see also \citealt{1994ApJ...420..213K}).},
internally isothermal, and in pressure equilibrium with the surrounding
medium. The unperturbed ambient medium is assumed to be isothermal (with
temperature $T_{\rm ism}=10^4$ K, corresponding to an isothermal
sound speed $c_{\rm s} = 11.5$ km s$^{-1}$) and homogeneous (with
hydrogen number density $n_{\rm ism} = 0.1$ cm$^{-3}$). The total
mass of the cloud is $\sim 0.13 ~M_{\rm sun}$. Table \ref{tab1} summarizes
the initial physical parameters characterizing the unperturbed ambient
medium and the spherical cloud. The shock propagates with a velocity $w=
\mach c_{\rm ism}$ in the ambient medium, where $\mach$ is the shock Mach
number, and $c_{\rm ism}$ is the sound speed in the interstellar medium;
the post-shock conditions of the ambient medium well before the impact
onto the cloud are given by the strong shock limit (\citealt{zel66},
see also Paper I). The fluid is assumed to be fully ionized, and is
regarded as a perfect gas (with a ratio of specific heats $\gamma = 5/3$).

\begin{table}
\caption{Summary of the initial physical parameters characterizing the
unperturbed ambient medium and the spherical cloud.}
\label{tab1}
\begin{tabular}{lccc}
\hline
\hline
 & Temperature~ & Density~~~ & Cloud radius \\
\hline
{\it ISM} & $10^4$ K & $0.1$ cm$^{-3}$ & - \\
{\it Cloud} & $10^3$ K & $1.0$ cm$^{-3}$ & 1 pc \\
\hline
\end{tabular}
\end{table}

The plasma evolution is derived by solving the time-dependent fluid
equations of mass, momentum, and energy conservation. We take into
account the thermal conduction (according to the formulation of
\citealt{spi62}), including the free-streaming limit (saturation)
on the heat flux (\citealt{cm77}, \citealt{1984ApJ...277..605G},
\citealt{1989ApJ...336..979B}, \citealt{2002A&A...392..735F}, and
references therein), and the radiative losses from an optically thin
plasma (e.g. \citealt{rs77}, \citealt{mgv85} and later upgrades).
Continuity equation of a tracer of the original cloud material is solved
in addition to our set of hydrodynamic equations. The calculations are
performed using the \FLASH\ code (\citealt{for00}) with customized
numerical modules that treat thermal conduction and optically thin
radiative losses (see Paper I).

 \begin{table*}[!ht]
 \caption{Parameters of the simulated shock-cloud interactions.}
 \label{tab2}
 \begin{tabular}{llccccccc}
 \hline
 \hline
 Run & Geometry & $\mach^a$ & $\chi^b$ & $w^c$ & $T_{\rm psh}^d$ & 
 $n_{\rm psh}^e$ & $\tau_{\rm cc}^f$ & 
 therm. cond. \\
 &  & & & [km s$^{-1}$] & [$10^6$ K] & [cm$^{-3}$] & [$10^3$ yr] & 
   \& rad. losses\\
 \hline
 HYm30c10 $^g$& 3-D cart. ($x,y,z$) & 30 & 10 & 344 & 1.7 & 0.4 & 9.1 & no \\
 HYm50c10     & 3-D cart. ($x,y,z$) & 50 & 10 & 574 & 4.7 & 0.4 & 5.4 & no \\
 RCm30c10     & 2-D cyl. ($r,z$)    & 30 & 10 & 344 & 1.7 & 0.4 & 9.1 & yes\\
 RCm50c10     & 2-D cyl. ($r,z$)    & 50 & 10 & 574 & 4.7 & 0.4 & 5.4 & yes\\
 \hline
 \\
 \multicolumn{5}{l}{$^a$ Shock Mach number.} &
 \multicolumn{4}{l}{$^e$ Density of the post-shock ambient medium.} \\
 \multicolumn{5}{l}{$^b$ Density contrast cloud / ambient medium.} &
 \multicolumn{4}{l}{$^f$ Cloud crushing time (\citealt{1994ApJ...420..213K}).}\\
 \multicolumn{5}{l}{$^c$ Velocity of the SNR shock.} &
 \multicolumn{4}{l}{$^g$ Run derived from HYm50c10 through Mach}\\
 \multicolumn{5}{l}{$^d$ Temperature of the post-shock ambient medium.} & 
 \multicolumn{4}{l}{~~ scaling (see Paper I).}\\
 \end{tabular}
 \end{table*}

To study the X-ray emission expected during the shock-cloud collisions,
simulated in Paper I (with an effective spatial resolution of $\approx
132$ zones per cloud radius) that allow us to explore two different
physical regimes in which either thermal conduction or radiative cooling
plays a dominant role. The set of simulations includes: two models
neglecting thermal conduction and radiation and considering the $\mach=30$
and $\mach=50$ shock cases in a 3-D cartesian coordinate system $(x,y,z)$
(runs HYm30c10 and HYm50c10, respectively\footnote{Run HYm30c10 has been
derived from run HYm50c10 through the scaling $t \rightarrow t\mach$,
$u\rightarrow u/\mach$, $T\rightarrow T/\mach^2$ (where $t$ is the time,
$u$ the gas velocity, and $T$ the temperature), with distance, density,
and pre-shock pressure left unchanged (the so-called Mach-scaling;
\citealt{1994ApJ...420..213K}, see also Paper I).}); two models with
thermal conduction and radiation, considering the $\mach=30$ and
$\mach=50$ shock cases, and in a 2-D cylindrical coordinate system
$(r, z)$ (runs RCm30c10 and RCm50c10, respectively). The models
neglecting both thermal conduction and radiation have been computed in a
3-D Cartesian coordinate system $(x,y,z)$, in order to describe accurately
the hydrodynamic instabilities developing at the boundaries of the shocked
cloud (e.g. \citealt{1995ApJ...454..172X}; see also Paper I). As we have
demonstrated in Paper I, however, the heat conduction rapidly damps the
hydrodynamic instabilities and, in this case, the essential evolutionary
features of the system can be adequately captured in a model using 2-D
cylindrical coordinate system $(r,z)$. In all the cases considered,
the cloud is initially 10 times denser than the surrounding medium
(hydrogen number density of the cloud, $n_{\rm cl} = 1$ cm$^{-3}$,
see Tab. \ref{tab1}). Table \ref{tab2} summarizes the physical
parameters characterizing the simulations, namely the shock Mach number,
$\mach$, the density contrast between the cloud and the ambient medium,
$\chi = n_{\rm cl}/n_{\rm ism}$, the velocity of the SNR shock, $w$, the
temperature and density of the post-shock ambient medium, $T_{\rm psh}$
and $n_{\rm psh}$ respectively, and the cloud crushing time, $\tau_{\rm
cc}$, i.e. the characteristic time for the transmitted shock to cross
the cloud (\citealt{1994ApJ...420..213K}, see also Paper I).

\subsection{Synthesis of the X-ray emission}
\label{sec3}

From the model results we synthesize the X-ray emission of the shock-cloud
system in different spectral bands of interest. The results of numerical
simulations are the evolution of temperature, density, and velocity of
the plasma in the spatial domain. In the case of 2-D simulations, we
reconstruct the 3-D spatial distribution of these physical quantities by
rotating the 2-D slab around the symmetry $z$ axis ($r=0$). The emission
measure in the $j$-th domain cell is $\em_{\rm j} = n_{\rm Hj}^2 V_{\rm
j}$ (where $n_{\rm Hj}^2$ is the hydrogen number density in the cell,
and $V_{\rm j}$ is the cell volume). We derive distributions of emission
measure vs. temperature, EM($T$), in selected regions by binning the
emission measure values in those regions into slots of temperature; the
range of temperature [$4 < \log T (\mbox{K}) < 7$] is divided into 75
bins, all equal on a logarithmic scale. From the EM($T$) distributions,
we synthesize the X-ray spectrum, using the MEKAL spectral synthesis code
(\citealt{mgv85}; \citealt{kaas92} and later upgrades), assuming solar
metal abundances (\citealt{1991sia..book.1227G}).

To derive spatial maps of the X-ray emission from the shock-cloud system,
we assume that the primary shock front propagates perpendicularly to
the line-of-sight (in the following LoS) and that the depth along the
LoS is 10 pc with inter-cloud conditions for the medium outside the
numerical spatial domain. The X-ray spectra integrated along the LoS
and on pixels of size corresponding to the spatial resolution of the
numerical simulations are then integrated in selected energy bands,
obtaining the X-ray images of the shock-cloud system.

\section{Results}
\label{sec4}

In Paper I, we have studied and discussed the hydrodynamics of the
shock-cloud interaction for the cases considered here. We found that
the shocked cloud evolves in cold, dense, and compact cooling-dominated
fragments surrounded by a hot diluted thermally conducting corona,
when the radiative losses are dominant ($\mach = 30$ shock case; see
Figs. 7 and 8 in Paper I). In this case, the radiative cooling strongly
modifies the structure of the shock transmitted into the cloud, leading
to a cold and dense gas phase. When the thermal conduction is the dominant
process ($\mach = 50$ shock case), the shocked cloud evaporates in a
few dynamical time-scales, and a transition region from the inner part
of the cloud to the ambient medium is generated (see Figs. 4 and 5 in
Paper I).

\subsection{Emission measure vs. temperature}
\label{emt}

\begin{figure*}[!t]
  \centering 
  \includegraphics[width=17cm]{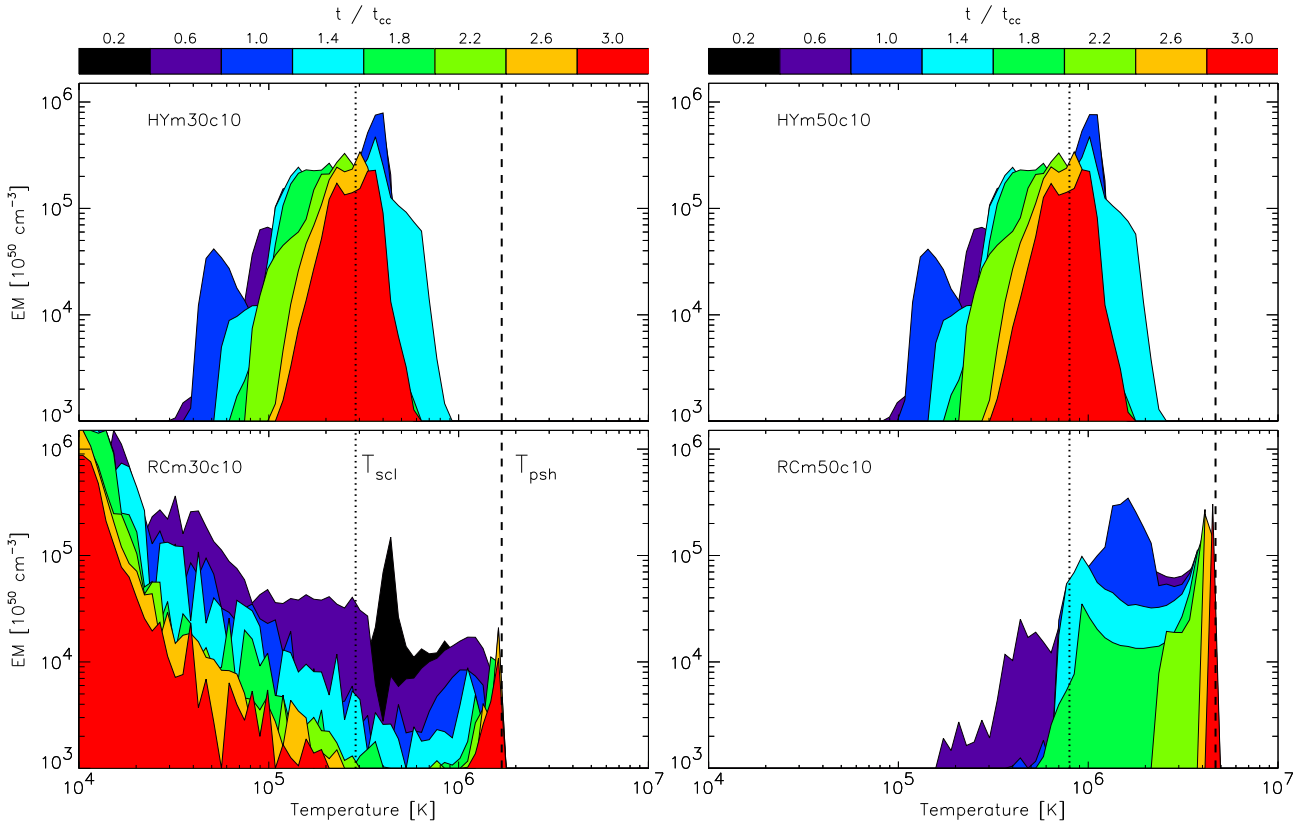}
  \caption{Evolution of the EM($T$) distributions of the cloud for
  $\mach=30$ (left panels) and $\mach=50$ (right panels) cases;
  upper panels show the result for models without thermal conduction
  and radiation (HY models), lower panels for models with both effects
  (RC models). The EM($T$) distributions are sampled every $0.4~\tau_{\rm
  cc}$ since $t=0.2~\tau_{\rm cc}$. The time sequence follows the color
  codes as reported in the color bar. The temperatures of the shocked
  ambient medium ($T_{\rm psh}$, see Tab. \ref{tab2}) and of the shocked
  cloud medium assuming negligible thermal conduction ($T_{\rm scl}$, see
  text) are marked with vertical dashed and dotted lines, respectively.}
\label{fig1} \end{figure*}

We use the cloud tracer mentioned in Sect. \ref{sec2} to identify zones
whose content is the original cloud material by more than 90\%.  From
these zones, we then derive the EM($T$) distribution of the cloud. Fig.
\ref{fig1} shows the cloud EM($T$) evolution for the $\mach=30$ (left
panels) and $\mach=50$ (right panels) cases, either without thermal
conduction and radiation (hereafter HY models; upper panels) or with
both effects (hereafter RC models; lower panels). We show the
EM($T$) distributions sampled at steps of $0.4~\tau_{\rm cc}$ since
$t=0.2~\tau_{\rm cc}$.

Fig. \ref{fig1} shows that, in HY models, the EM($T$) distribution
of the cloud is steadily centered around the temperature of the shock
transmitted into the cloud, $T_{\rm scl} \approx \beta T_{\rm psh}/\chi$,
where $\beta\approx 1.7$ (see Paper I), $\chi = 10$; we obtain $T_{\rm
scl} \approx 0.3$ MK for $\mach = 30$ and $T_{\rm scl} \approx 0.8$ MK
for $\mach = 50$ (see dotted lines in Fig. \ref{fig1}). The evolution
of the EM($T$) distribution for $\mach=30$ and for $\mach=50$ cases is
similar, according to the Mach-scaling (\citealt{1994ApJ...420..213K},
Paper I): it rapidly becomes quite broad, covering more than a decade in
temperature around $T_{\rm scl}$; then, at late stages, it gets narrower.

The EM($T$) distribution obtained from RC models significantly changes,
depending on which process is dominant. When the radiative losses dominate
(RCm30c10; lower left panel in Fig. \ref{fig1}), the EM($T$) distribution
below 1 MK evolves toward a steep power law (with negative index),
drifting to the cold side due to the progressive cooling of the plasma.
A small fraction of cloud material gradually thermalizes by conduction
to the temperature of the surrounding medium, forming a small peak
centered at $T_{\rm psh}$. Thus, at variance from pure hydrodynamics,
the plasma splits into two separate thermal components: a cold dense core
($T< 0.1$ MK, see Paper I) and a hot diluted corona ($T\approx T_{\rm psh}
= 1.7$ MK).

In the conduction-dominated Mach 50 case (RCm50c10; lower right panel in
Fig. \ref{fig1}), the EM($T$) distribution is initially broad and centered
at the temperature of the shock transmitted into the cloud, $T_{\rm scl}$;
then its maximum gradually shifts to higher and higher temperatures up to
$T_{\rm psh}\sim 4.7$ MK getting more peaked, due to the thermalization
of the cloud material to $T_{\rm psh}$.

\subsection{X-ray emission}
\label{emission}

\begin{figure}[!t]
  \centering
  \includegraphics[width=8.5cm]{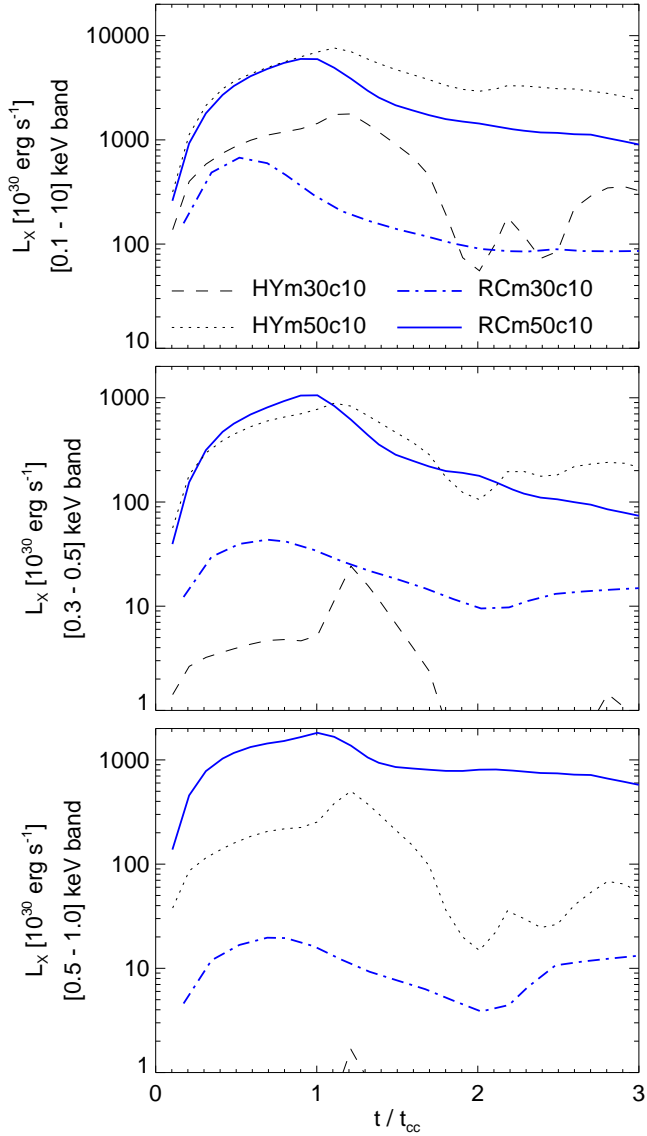}
  \caption{X-ray light curve of the cloud in the  $[0.1-10]$ keV
  (top panel), $[0.3-0.5]$ keV (middle panel), and $[0.5-1.0]$ keV
  (bottom panel) bands of the shocked cloud derived for the $\mach =30$
  and $\mach =50$ shock cases with (RC models) or without (HY models)
  thermal conduction and radiative cooling.}
\label{fig2} \end{figure}

Fig. \ref{fig2} shows the cloud X-ray light curves in the broad $[0.1-10]$
keV band and in the $[0.3-0.5]$ keV, and $[0.5-1.0]$ keV bands typically
selected for the analysis of evolved SNR shock-cloud interactions (see,
for instance, \citealt{2005A&A...442..513M}). The figure shows the X-ray
luminosity, $L_{\rm X}$, of the shocked cloud only and does not consider
the contribution originating from the shocked ambient medium surrounding
the cloud. As expected, $L_{\rm X}$ is larger in the hotter $\mach =50$
case than in the $\mach =30$ case in all the energy bands and in both HY
and RC models. In all the cases, the X-ray luminosity of the shocked cloud
reaches its maximum quite early, around $t\sim \tau_{\rm cc}$, and then
decreases (even by one order of magnitude, for instance, in HYm50c10 in
the $[0.3-0.5]$ keV band); therefore, in the X-ray band, shocked
interstellar gas clouds will be preferentially visible during the early
phases of the shock-cloud collision.

\begin{figure}[!t]
  \centering
  \includegraphics[width=8.5cm]{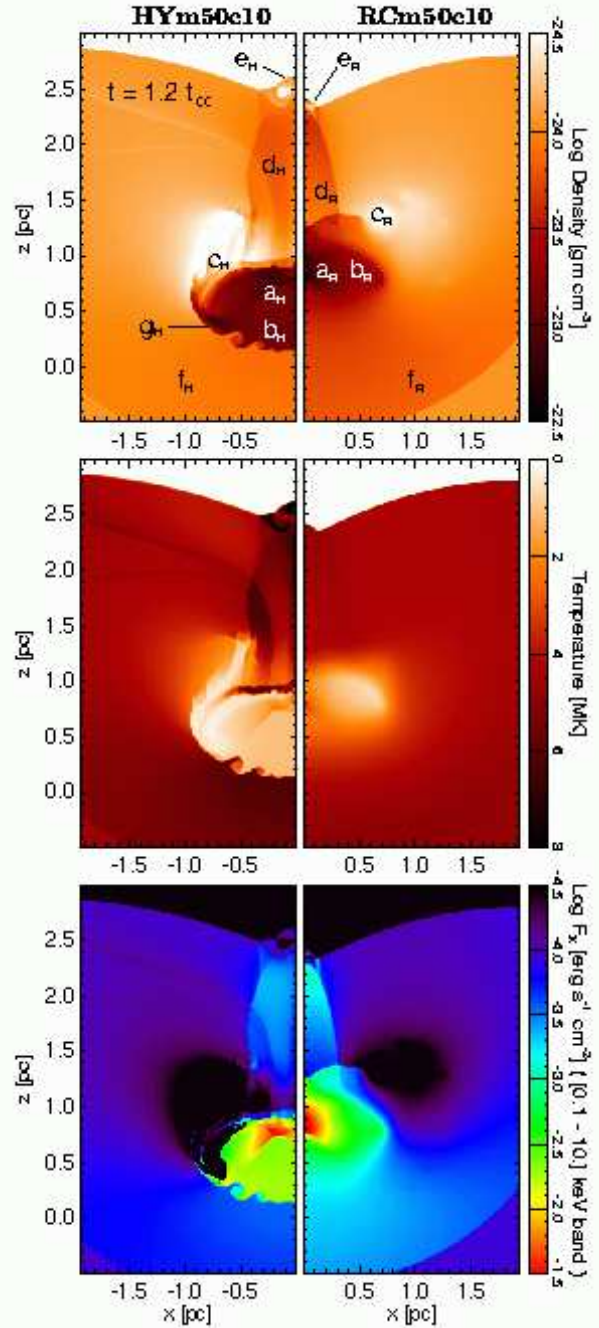}
  \caption{2-D sections in the $(x,z)$ plane of the mass density
   distribution (gm cm$^{-3}$; top panels) in log scale, temperature (MK;
   middle panels), and X-ray emission in the $[0.1-10]$ keV band (erg
   s$^{-1}$; bottom panels), in log scale, derived from runs HYm50c10
   (left panels) and RCm50c10 (right panels) at $t=1.2~ \tau_{\rm
   cc}$. Labeled regions in the upper panels are discussed in the text.}
\label{fig3}
\end{figure}

We now discuss in detail the evolution of the cloud morphology as detected
in the X-ray band. We expect the richest scenario from the hottest
$\mach=50$ case. Fig. \ref{fig3} shows 2-D sections in the $(x,z)$
plane of the mass density distribution, $\rho$, the temperature, $T$,
and the X-ray flux in the $[0.1-10]$ keV band, $F_{\rm X}$, derived
from HYm50c10 and RCm50c10 models at $t=1.2~ \tau_{\rm cc}$, just after
the maximum luminosity of the cloud (see Fig. \ref{fig2}). The maps of
$\rho$ and $F_{\rm X}$ are in log scale to highlight structures with
very different density and X-ray fluxes.

At this stage of evolution, the whole cloud material has already been
shocked. The size of the cloud ($\lsim 1$ pc) is smaller than that of
the original unshocked cloud ($2~r_{\rm cl} = 2$ pc) due to the 
cloud compression. The core of the cloud is a high density region
($a_{\rm H}$ and $a_{\rm R}$, see upper panel in Fig. \ref{fig3}) where
primary and reverse shocks transmitted into the cloud are colliding (see
also Paper I). A low density region ($c_{\rm H}$ and $c_{\rm R}$) due
to a large vortex ring has developed just behind the cloud. In HYm50c10,
hydrodynamic instabilities are developing at the cloud boundaries: the
combined effect of instabilities and shocks transmitted into the cloud
leads to unstable high-density regions at the cloud boundaries ($g_{\rm
H}$). In RCm50c10, the thermal conduction suppresses the instabilities
and leads to smooth gradients of density and temperature from the
inner part of the cloud to the ambient medium. In both cases, the global
forward shock has converged on the symmetry axis ($z$-axis), and undergoes
a conical self-reflection, forming the primary Mach reflected shocks
($d_{\rm H}$ and $d_{\rm R}$) and the stem bulge at the base of the
secondary vortex sheets near the symmetry axis ($e_{\rm H}$ and $e_{\rm
R}$; see Fig. 6 in \citealt{2002ApJ...576..832P} for a detailed
description of the flow structures developing during the shock-cloud
interaction). The reflected bow shock is visible as a curved region
extending into the shocked ISM right below and along the sides of the
cloud (regions $f_{\rm H}$ and $f_{\rm R}$).

The comparison between upper and lower panels of Fig. \ref{fig3} shows
that, in both models, the region with the highest X-ray flux is in the
core of the cloud (regions $a_{\rm H}$ and $a_{\rm R}$). In HYm50c10,
the X-ray image of the shocked cloud has a very sharp boundary and even
the hydrodynamic instabilities at the cloud boundary are clearly marked
in the X-rays (region $b_{\rm H}$); high X-ray flux also originates
from the unstable high-density regions at the cloud boundaries (region
$g_{\rm H}$). In RCm50c10, instead, the emission from region $b_{\rm
R}$ is more diffuse, varying smoothly in the radial direction from the
center of the cloud. Fig. \ref{fig3} also shows that the X-ray emission
density in the reflected bow shock (regions $f_{\rm H}$ and $f_{\rm R}$)
and in the primary Mach reflected shocks ($d_{\rm H}$ and $d_{\rm R}$)
is slightly higher (by a factor $\sim 2$ in HYm50c10, and by a factor
$\sim 4$ in RCm50c10) than that of the post-shock ambient medium not
involved in the shock-cloud interaction. On the other hand, the low
density region ($c_{\rm H}$ and $c_{\rm R}$) and the stem bulge ($e_{\rm
H}$ and $e_{\rm R}$) are characterized by very low X-ray emission.

\begin{figure}[!t]
  \centering
  \includegraphics[width=8cm]{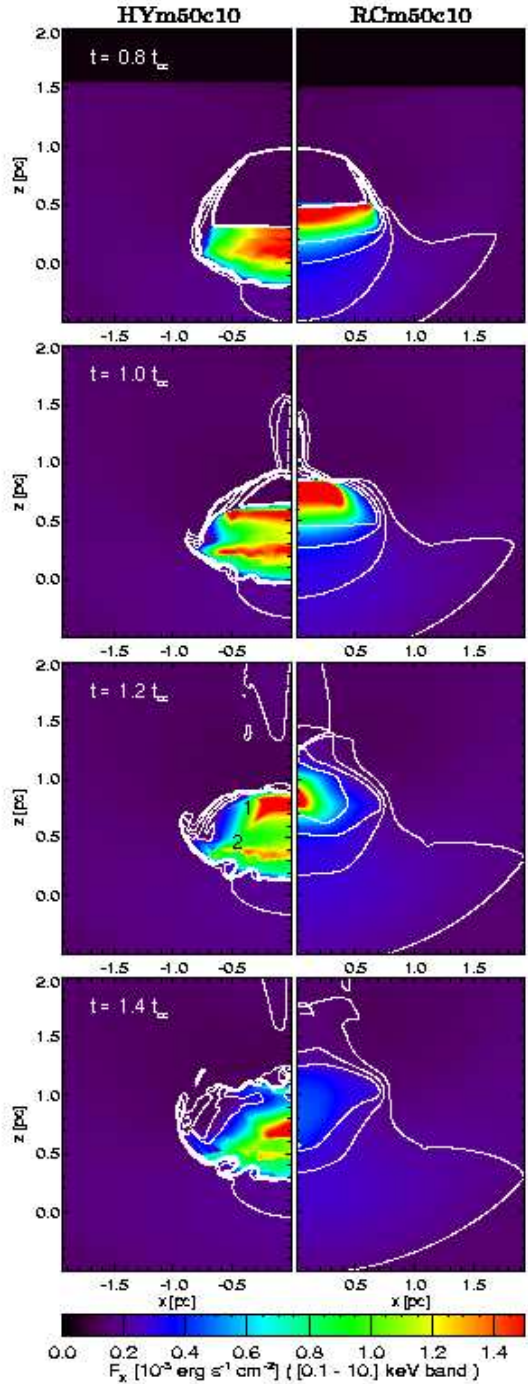}
  \caption{X-ray images in the [$0.1-10$] keV band in linear scale
  derived from models HYm50c10 (left panels) and RCm50c10 (right panels)
  at the labeled times. Contours are the mass density distribution in
  the $(x,z)$ plane, corresponding to $\log \rho~ \mbox{(gm cm$^{-3}$)} =
  -23.9,~ -23.7,~ -23.5,~ -23.3$. Regions ``1'' and ``2'' in the third
  row originate from regions $a_{\rm H}$ and $g_{\rm H}$, respectively,
  shown in Fig. \ref{fig3}.}
\label{fig4} \end{figure}

Fig. \ref{fig4} shows the map of X-ray emission in the $[0.1-10]$
keV band (in linear scale) integrated over 10 pc along the LoS (see
Sect. \ref{sec3}) for HYm50c10 and RCm50c10 at four selected epochs around
the time of the maximum cloud X-ray luminosity (see Fig. \ref{fig2}). The
superimposed contours are the mass density distribution (in log scale)
in the $(x,z)$ plane. The third row of plots corresponds to the time
of Fig. \ref{fig3} ($t=1.2~\tau_{cc}$). The highest integrated emission
originates in the core of the shocked cloud in both models; two separate
high-emission regions are visible in HYm50c10 around $t=\tau_{\rm cc}$:
the upper region (labeled ``1'' in the third row of Fig. \ref{fig4}) is
the high-density region $a_{\rm H}$ discussed above (see Fig. \ref{fig3}),
whereas the lower one (region labeled ``2'') originates from the
integration along the LoS of the emission of high-density unstable
regions at the cloud boundary (region $g_{\rm H}$ in the upper panel in
Fig. \ref{fig3}). In HYm50c10, the hydrodynamic instabilities are
no longer clearly distinguishable after integration along the LoS. In
RCm50c10, the brightest portion of the shocked cloud corresponds to
the high-density region ($a_{\rm R}$), appears significantly smaller
and shorter-lived than in HYm50c10 and, in general, the cloud surface
brightness rapidly approaches the values of the surrounding medium.
The reflected bow shock (regions $f_{\rm H}$ and $f_{\rm R}$) and the
primary Mach reflected shocks (regions $d_{\rm H}$ and $d_{\rm R}$)
have a surface brightness more than a decade lower than that of the
cloud core.

\begin{figure}[!t]
  \centering
  \includegraphics[width=8.cm]{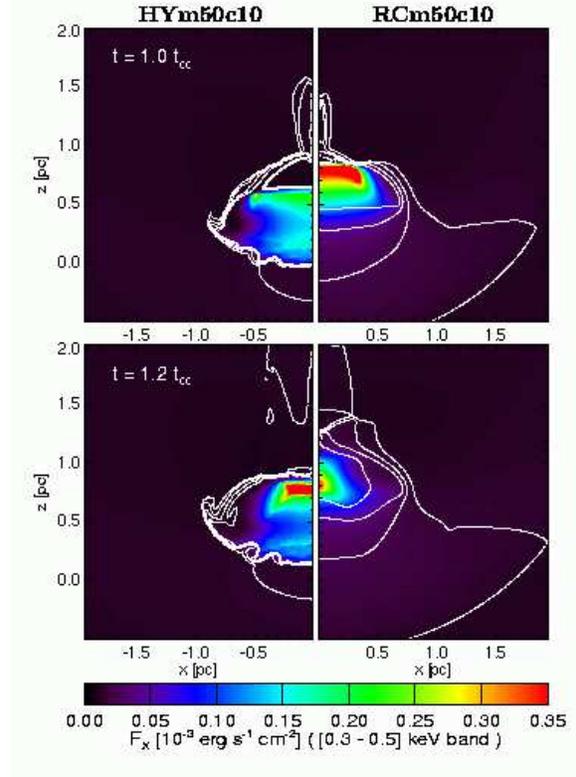}
  \caption{As in Fig. \ref{fig4} for the two labeled times and for the
  [$0.3-0.5$] keV band.}
\label{fig5} \end{figure}

Fig. \ref{fig5} shows that the highest emission comes from the core of the
shocked cloud (regions $a_{\rm H}$ and $a_{\rm R}$) in the $[0.3-0.5]$ keV
band, and is maximum at $t\approx \tau_{\rm cc}$ for RCm50c10 and at
$\approx 1.2~\tau_{\rm cc}$ for HYm50c10. In model RCm50c10, the cloud
fades out earlier than in HYm50c10 because of the dissipation by thermal
conduction.

\begin{figure}[!t]
  \centering
  \includegraphics[width=8.cm]{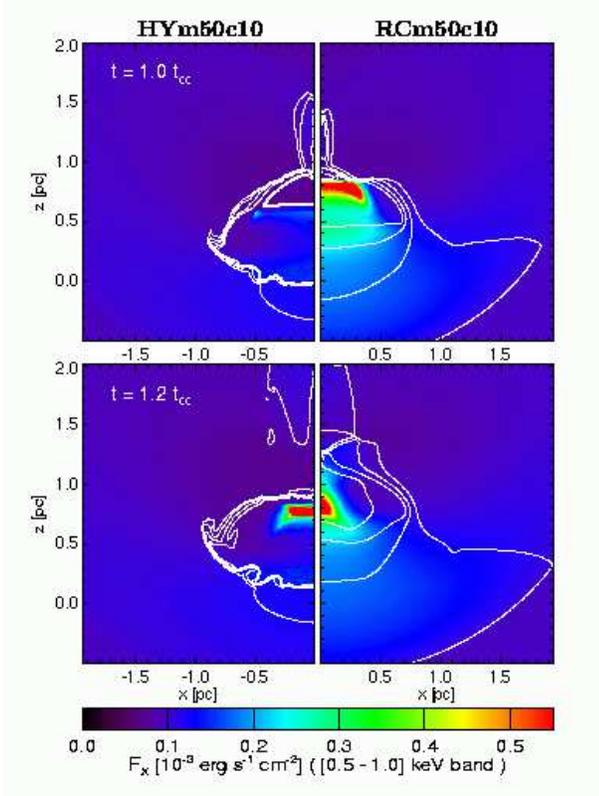}
  \caption{As in Fig. \ref{fig5} for the [$0.5-1.0$] keV band.}
\label{fig6}
\end{figure}

In the higher energy $[0.5-1.0]$ keV band, shown in Fig. \ref{fig6},
the shocked cloud is bright only around $\tau_{\rm cc}$. In particular,
in HYm50c10, a small fraction of the cloud (the high-density region
$a_{\rm H}$; see Fig. \ref{fig3}) has significant emission only for a
very short time around $1.2~ \tau_{\rm cc}$ (see Fig. \ref{fig6}). In
RCm50c10, instead, the X-ray image of the cloud appears more extended
and diffuse than in HYm50c10, and the cloud is already visible at
$t=0.2~\tau_{\rm cc}$ (not shown in Fig. \ref{fig6}) and remains bright
until $t=1.4~\tau_{\rm cc}$. This larger-lasting emission is due to the
increased cloud X-ray emission at higher energies, determined by the
thermal conduction that heats the cloud material to higher temperatures.

For further details on the evolution of the X-ray emission in the three
bands selected, see the on-line material.

\subsection{Median energy of X-ray photons}

The map of the median energy of X-ray photons (hereafter the MPE map)
is a practical tool to convey at the same time both spatial and spectral
information on the emitting plasma at high resolution
(\citealt{2004ApJ...614..508H}; see also \citealt{2005A&A...442..513M}).

\begin{figure}[!t]
  \centering
  \includegraphics[width=8.cm]{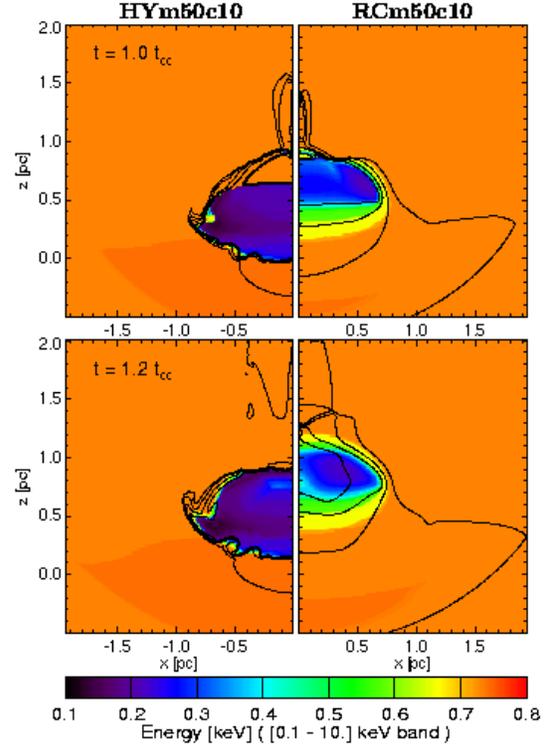}
  \caption{Median photons energy maps in the $[0.1-10]$ keV band
   derived for HYm50c10 (left panels) and RCm50c10 (right panels) at
   the labeled times. Contour plots as in Fig. \ref{fig4}.}
\label{fig7}
\end{figure}

\begin{figure}[!t]
  \centering
  \includegraphics[width=8.cm]{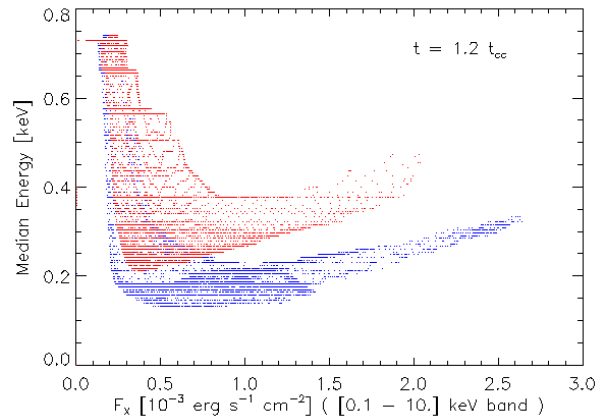}
  \caption{Median photon energy, $\ener$, versus X-ray flux in the
   $[0.1-10]$ keV band, $F_{\rm X}$, scatter plot derived for HYm50c10
   (blue) and RCm50c10 (red) at $t = 1.2 ~\tau_{\rm cc}$.}
\label{fig8}
\end{figure}

Fig. \ref{fig7} shows the model MPE maps at the times of Figs. \ref{fig5}
and \ref{fig6} obtained from the spectra synthesized in the [$0.1-10$]
keV band (see Sect. \ref{sec3}). By comparing the MPE maps in Fig.
\ref{fig7} with the X-ray images in Fig. \ref{fig4}, we note that,
during the whole evolution, the X-ray emission is high where the median
photon energy $\ener$ is low. This result is evident in the $\ener$
versus $F_{\rm X}$ scatter plot derived for both HYm50c10 and RCm50c10
models at $t = 1.2 ~\tau_{\rm cc}$ (Fig. \ref{fig8}). In fact, most of
the brightest pixels are in the shocked cloud, i.e. plasma with
temperature lower than that of the shocked surrounding medium. Fig.
\ref{fig8} also shows that, for $F_{\rm X} > 0.5\times 10^{-3}$ erg
s$^{-1}$ cm$^{-2}$, $\ener$ increases with $F_{\rm X}$ indicating that
the cloud plasma is far away from pressure equilibrium.

Fig. \ref{fig7} also shows that, in HYm50c10, the low $\ener$ region
is rather uniform with $\ener \approx 0.2$ keV, and its boundaries are
sharp. In model RCm50c10, instead, the thermal conduction smoothes the
energy gradient: in the low $\ener$ region, $\ener$ increases smoothly
from the cloud center to the surrounding medium. The minimum $\ener$
value is higher ($\sim 0.3$ keV) than that in HYm50c10 ($\sim 0.2$ keV)
because of the heat conducted into the cloud.

\section{Discussion and conclusion}
\label{sec5}

We derived the X-ray emission predicted by hydrodynamic modeling of
the interaction of a SNR shock wave with an interstellar gas cloud. Our
forward modeling allows us to link model results to observable quantities
and to investigate the observability of features predicted by models.

Our analysis has shown that the morphology of the X-ray emitting
structures is significantly different from the morphology of the flow
structures originating from the shock-cloud interaction. For instance,
the complex pattern of shocks (e.g. external reverse bow shock, shocks
transmitted into the cloud, Mach reflected shocks at the symmetry axis,
etc.) as well as other flow structures (e.g. hydrodynamic instabilities,
the stem bulge at the base of the secondary vortex sheets near the
symmetry axis, etc.) caused by the shock-cloud collision are visible
in the density maps, but they are never clearly distinguishable in
X-ray images (cf. upper panel in Fig. \ref{fig3} and third row of
Fig. \ref{fig4} at $t=1.2~\tau_{cc}$). Indeed, the morphology of the X-ray
emitting structures appears quite simple in all the cases examined. The
largest contribution to the X-ray emission originates from the core of
the cloud where primary and reverse shocks transmitted into the cloud
collide. The bright core is surrounded by a diffuse and faint region
associated with the outer portion of the cloud. The X-ray emission varies
smoothly in the radial direction from the bright core to the surrounding
medium.

The hydrodynamic instabilities, developing at the cloud boundaries in
models without thermal conduction, are never clearly visible in the X-ray
band because faint and washed out by integration along the LoS. On the
other hand, the interaction of the instabilities with shocks transmitted
into the cloud produces a bright region with luminosity comparable to
that of the cloud core (region labeled ``2'' in Fig. \ref{fig4}). At
variance with models including thermal conduction, therefore, in HY models
two separate bright regions develop inside the shocked cloud. This has
an important implication on the diagnostics. In fact, in Paper I, we
have shown that the thermal conduction is very effective in suppressing
hydrodynamic instabilities: the evidence of these instabilities during
the shock-cloud interaction would be an indication that the thermal
conduction is strongly inhibited (for instance by an ambient magnetic
field). Our analysis points out that, unfortunately, the X-ray band cannot
give strong indications about hydrodynamic instabilities in any case.

Our modeling has also shown that shocked interstellar gas clouds reach
their maximum X-ray luminosity around $t\sim \tau_{\rm cc}$. The size
of the bright region in X-ray maps varies during the shock-cloud
interaction: the maximum extension is reached at epochs $< \tau_{\rm cc}$
and is always significantly smaller than the original cloud diameter. The
light curve of the shocked cloud and the evolution of the bright region
indicate that shocked clouds are expected to be preferentially observed
in the X-rays during the early phases of shock-cloud collision.

As an example, in the $\mach = 50$ shock case considered here, the
shocked cloud has total luminosity in the $[0.5-1.0]$ keV band $L_{\rm X}
\gsim 10^{33}$ erg s$^{-1}$ during the period $0.4~\tau_{\rm cc} < t <
1.3~\tau_{\rm cc}$ (see Fig. \ref{fig2}). Since the XMM-Newton/EPIC-MOS
sensitivity limit in the $[0.5-2.0]$ keV band is $F_{\rm epic} \approx
10^{-14}$ erg cm$^{-2}$ s$^{-1}$ for an exposure time of $10^4$ s
(\citealt{2001A&A...365L..51W}), our analysis suggests that such emission
should be detectable as far as $\approx 30$ kpc.

The modeling shows that thermal conduction and radiative cooling can
lead to two different gas phases emitting in different energy bands:
a cooling dominated core which ultimately fragments into cold, dense,
and compact filaments emitting at low energies (e.g. optical band), and
a hot thermally conducting corona emitting at high energies (e.g. soft
X-rays). Both phases are clearly present in the $\mach=30$ case but only
the hot one in the $\mach=50$ case because thermal conduction is highly
effective. As an implication, we expect that the X-ray emission morphology
and spectrum of the bright cloud region should be sensitive to thermal
conduction effects. In fact, thermal conduction makes the X-ray bright
region smaller, more diffuse, and shorter-lived than that expected when
thermal conduction is neglected. Also, we found that the median photons
energy of the bright region is higher in models with thermal conduction.
As a final diagnostic consideration, we note that observing smooth
gradients of emission and median photon energy would indicate that the
thermal conduction is efficient.

The results presented here illustrate the X-ray radiation emitted during
the shock-cloud collision. Our analysis provides a way: 1) to link the
features expected to emit X-rays with plasma structures originating during
the shock-cloud collision, and 2) to investigate the effects on the X-ray
emission of the different physical processes at work. These results will
be a guide for the interpretation of X-ray observations of middle-aged
X-ray SNR shells whose morphology is affected by ISM inhomogeneities
(e.g. the Cygnus Loop, the Vela SNR, G272.2-3.2, etc.). However, a
more direct comparison of model results with supernova remnant X-ray
observations requires to include instrumental response and sensitivity and
ISM absorption. In a companion paper (Orlando et al., in preparation),
we will step forward to investigate in detail the direct diagnostics
and comparison with the data collected with the latest X-ray instruments
(i.e. Chandra, XMM-Newton).

\bigskip
\acknowledgements{We thank the referee for the useful comments and
suggestions. The \FLASH\ code and related software used in this work
was in part developed by the DOE-supported ASC / Alliance Center for
Astrophysical Thermonuclear Flashes at the University of Chicago.
The simulations have been executed on the IBM/Sp4 machine at CINECA
(Bologna, Italy), in the framework of the INAF-CINECA agreement on ``High
Performance Computing resources for Astronomy and Astrophysics', and
on the Compaq cluster at the SCAN facility of the INAF - Osservatorio
Astronomico di Palermo. The work of T.P. was supported by the US
Department of Energy under Grant No. B523820 to the Center of
Astrophysical Thermonuclear Flashes at the University of Chicago. This
work was supported by Ministero dell'Istruzione, dell'Universit\`a e
della Ricerca, by Istituto Nazionale di Astrofisica, and by Agenzia
Spaziale Italiana.}

\bibliographystyle{aa}
\bibliography{references}

\end{document}